\documentclass[twocolumn,superscriptaddress,showpacs,amsmath,amssymb,aps,prb]{revtex4-1}
\usepackage{graphicx}  
\usepackage{dcolumn}   
\usepackage{bm}        
\usepackage{amsmath}
\usepackage{mathtools}
\usepackage{amssymb}   
\usepackage{mathrsfs} 
\usepackage[colorlinks=True,urlcolor=blue,linkcolor=blue,citecolor=blue]{hyperref}
\usepackage{xcolor}
\usepackage{multirow}
\usepackage{setspace}
\usepackage{CJKutf8}
\usepackage{color,soul}
\usepackage{subfigure}
\usepackage[utf8]{inputenc}
\usepackage[T1]{fontenc}
\usepackage{lineno}
\usepackage{lmodern}
\usepackage[toc,page]{appendix}
\usepackage[markup=underlined]{changes}
\definechangesauthor[name={WQ Chen}, color=red]{wqc}

\begin{document}
\title{Fate of entanglement in one-dimensional fermion liquid with coherent particle loss}
\author{Wei-Zhu Yi}
\affiliation{Shenzhen Key Laboratory of Advanced Quantum Functional Materials and Devices, Southern University of Science and Technology, Shenzhen 518055,  China}
\affiliation{Department of Physics and Institute for Quantum Science and Engineering, Southern University of Science and Technology, Shenzhen 518055, China}
\author{ Hao-Jie Lin}
\affiliation{Department of Physics and Institute for Quantum Science and Engineering, Southern University of Science and Technology, Shenzhen 518055, China}
\author{Ze-Xun Lin}
\affiliation{Department of Physics, The University of Texas at Austin, Austin,
TX 78712, USA }
\affiliation{Department of Physics, Northeastern University, Boston, MA 02115,
USA}
\author{Wei-Qiang Chen}
\email{chenwq@sustech.edu.cn}

\affiliation{Shenzhen Key Laboratory of Advanced Quantum Functional Materials and Devices, Southern University of Science and Technology, Shenzhen 518055,  China}
\affiliation{Department of Physics and Institute for Quantum Science and Engineering, Southern University of Science and Technology, Shenzhen 518055, China}
\affiliation{International Quantum Academy, and Shenzhen Branch, Hefei National Laboratory, Futian District, Shenzhen, P. R. China}

\begin{abstract}
Quantum many-body systems and quantum devices experience the detrimental effects of noise and particle losses, necessitating their treatment as open quantum systems or, in approximation, as non-Hermitian systems. These systems exhibit nontrivial characteristics in their time evolution that differ significantly from closed systems. In this Letter, we study the dynamic properties of a one-dimensional fermionic system with adjacent-lattice particle loss. By utilizing time-dependent correlation matrix methods and bosonization techniques, we  demonstrate that, as the system evolves over time, its (bipartite) von Neumann entropy exhibits a universal behavior of rapid increase due to thermalization effects at short times, independent of the effective Hamiltonian and Liouvillian spectra, even in the presence of interactions. Additionally, we show that the asymmetric non-Hermitian terms in the effective Hamiltonian caused by adjacent-lattice quantum jumps lead to left-right asymmetry of quasiparticles in momentum space, which is ubiquitous in non-Hermitian skin effects and introduces momentum-space entanglement independent of the interaction strength at early times. Our study illuminates the universal fate of non-Hermitian fermionic liquids in the open quantum context, enriching our understanding of non-Hermitian many-body systems over the entire time range. Furthermore, our findings provide valuable insights for near-term quantum devices and the quantum simulation of open systems.
\end{abstract}
\maketitle
{\color{blue}\emph{Introduction.}}--
Open quantum system and non-Hermitian physics have become an increasingly attractive topic for research
. In recent years, there are lots of important findings in this area,
both theoretically\cite{li_topological_2020,pan_non-hermitian_2020,altland_symmetry_2021,lieu_tenfold_2020,yu_unsupervised_2021,gong_topological_2018,harari_topological_2018,ashida_non-hermitian_2020,PhysRevX.9.041015,PhysRevLett.80.5243,PhysRevLett.77.570} and experimentally\cite{hu_non-hermitian_2021,wang2021topological,bandres_topological_2018,xiao_non-hermitian_2020,zhao_non-hermitian_2019,zhang_observation_2021,chen_quantum_2021,zhu_simultaneous_2018,ozawa_topological_2019,doi:10.1126/science.aaw8205}. Especially,
in condensed matter physics, there have been lots of studies focusing
on non-Hermitian skin effect\cite{lee_anomalous_2016,li_critical_2020,yao_edge_2018,sun_geometric_2021,lee_hybrid_2019,borgnia_non-hermitian_2020,martinez_alvarez_non-hermitian_2018,song_non-hermitian_2019,okuma_topological_2020}, symmetry and topology of non-Hermitian
quantum phases\cite{lee_anomalous_2016,kawabata_classification_2019,leykam_edge_2017,yao_edge_2018,ding_emergence_2016,bergholtz_exceptional_2021,helbig_generalized_2020,luo_higher-order_2019,yang_jones_2020,borgnia_non-hermitian_2020,xue_non-hermitian_2020,song_non-hermitian_2019-1,song_p_2020,zhou_periodic_2019,liu_second-order_2019,kawabata_topological_2019,liang_topological_2013,sato_time-reversal_2012}, the phase transition and quantum criticality\cite{yin_kibble-zurek_2017,zeuner_observation_2015,ge_parity-time_2014,ashida_quantum_2016,arouca_unconventional_2020,longhi_topological_2019,okuma_topological_2019}, and the exceptional points in non-Hermitian system \cite{doi:10.1126/science.aar7709,ding_emergence_2016,denner_exceptional_2021,bergholtz_exceptional_2021,martinez_alvarez_non-hermitian_2018,molina_surface_2018}. However, most of them
are based on
the non-Hermitian band theory and short-time dynamics \cite{yokomizo_non-bloch_2019,xiao_non-hermitian_2020,mcdonald_phase-dependent_2018,rudner_topological_2009,gong_topological_2018,lee_topological_2019,shen_topological_2018}, which derives from a single-particle non-Hermitian Hamiltonian. 
Though the effective non-Hermitian Hamiltonian can describe certain effects in many cases, such as skin effects and quasi-particle lifetimes, it fails to capture the long time features of the open system due to its limitations. Therefore, there is a need to investigate non-Hermitian problems considering their fate in the long-term evolution, specifically in the context of many-body systems. Generally speaking, the non-hermicity of the effective Hamiltonian of an open system comes from the coupling
between the system and its environment.  It can be contained in the self-energy of single-particle Green's function resulting from the interactions, or equivalently it can be induced
by coupling to a Markovian environment while neglecting quantum
jumps or by postselection\cite{kozii_non-hermitian_2017,michishita_equivalence_2020}. 
The Schrödinger equation, with a non-Hermitian Hamiltonian, is often employed as an effective short-term approximation for observed systems. However, when considering these systems embedded within larger open systems, the long-term dynamics is more accurately captured by the Lindblad master equation \cite{lindblad1976generators,gorini_completely_1976},
\begin{align}
\label{M}
 \dot{\rho}=\mathcal{L}[\rho]:=\ensuremath{-\mathrm{i}[H_0,\rho]+\sum_{i}\left(L_{i}\rho L_{i}^{\dagger}-\frac{1}{2}\left\{ L_{i}^{\dagger}L_{i},\rho\right\} \right)}   
\end{align}
where $\rho$ is the density matrix of the system, $L_{i}$s are jump operators originating from the coupling between
the system and the environment. The effective non-Hermitian
Hamiltonian expresses as $H_{\text{eff }}=H_0-\frac{i}{2}\sum_{i}L_{i}^{\dagger}L_{i}$ by ignoring $L_{i}\rho L_{i}^{\dagger}$ terms.

 Recent progress in quantum information provides some novel tools to study quantum many-body systems, among that the von Neumann entropy (the entanglement entropy) is the most powerful in detecting system's universal properties, which is widely applied in multifarious fields\cite{RevModPhys.80.517,RevModPhys.82.277,calabrese_entanglement_2004}. In the case of zero-temperature
gapped quantum system, the entanglement entropy follows the area law\cite{brandao_area_2013,wolf_area_2008,RevModPhys.82.277}. For critical systems, entanglement entropy has logarithmic scaling
behavior\cite{swingle_entanglement_2010,vidal_entanglement_2003,RevModPhys.82.277}.
Under temporal evolution, entropy also tracks the universal dynamical properties of correlations in various systems. In ergodic systems, the time evolution follows a volume law of $t$. On the other hand, many-body localization exhibits a logarithmic behavior of $\text{ln}t$. The pattern of entropy-time dependence can be diverse in other problems such as unitary\cite{dora_crossover_2011,cazalilla_effect_2006,fagotti_evolution_2008,calabrese_time_2006,bardarson_unbounded_2012,serbyn_universal_2013} or non-unitary quantum quench\cite{moca_universal_2021,dora_quantum_2020,roberts2022fidelity},
generic dissipating\cite{bacsi_dissipation-induced_2020,bacsi_dynamics_2021,carollo_emergent_2021,bacsi_vaporization_2020} and generic stochastic dynamics \cite{alberton_entanglement_2021,ptaszynski_entropy_2019,okuma_quantum_2021,alba_spreading_2021,ashida_thermalization_2018,PhysRevB.98.205136,bensa_fastest_2021,nahum_measurement_2021,skinner_measurement-induced_2019}  in isolated, time-periodic
driven system\cite{apollaro_entanglement_2016,maskara_discrete_2021,berdanier_floquet_2017,ponte_many-body_2015}, or non-equilibrium systems\cite{alba_noninteracting_2021,breuer_colloquium_2016,calabrese_entanglement_2018,bensa_fastest_2021}.

In this letter, we study a generalized dimerized model whose coupling to
the environment is described by dissipative quantum-jump terms. It is observed that the effective Hamiltonian shares particular information with the Liouvillian in terms of eigenspectrum. By calculating the time-dependent correlation function, we demonstrate that the long-term behavior of the system is closely linked to the effective non-Hermitian Hamiltonian, even though it is not governed by it. Using the time-dependent correlation matrix method, we obtain the time-dependent entropy, which exhibits universal features in its time evolution. We show that when the quantum jumps are turned on, the system\textquoteright s
entropy surges in a short time signaling it quickly thermalized. The entropy then decreases, which manifests the
quantum dissipative effect. The decaying behavior depends not on whether the effective Hamiltonian is gapped or not, but dominantly on whether the Liouvillian spectra of the system is gapped or gapless. Next, we turn our attention to to momentum space and observe an asymmetry in the dynamics of left and right movers, which serves as a reminiscent of the non-Hermitian skin effect (NHSE). In addition, we provide a detailed discussion of the system with an included interaction, analyzed using the method of bosonization. We find that there is an interaction-independent momentum-space entanglement between bosonic quasi-particles, which arises from the left-right asymmetry. Overall, we believe our work serves as a guideline for future study of many-body problems in open quantum systems and dynamical non-Hermitian systems.

\begin{figure*}
\includegraphics[scale=1.3]{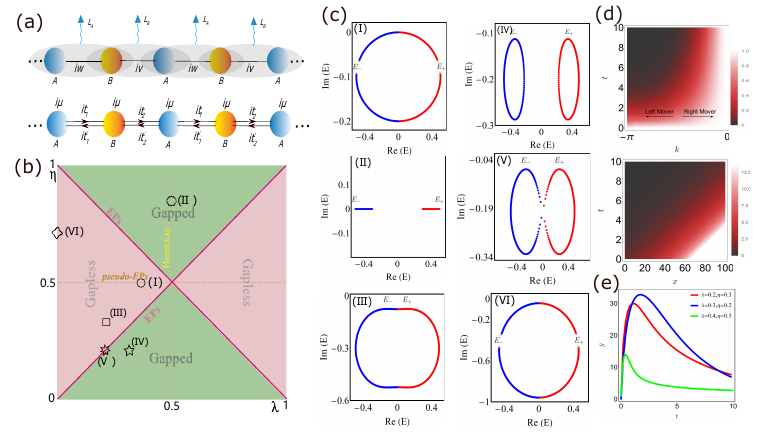}
\caption{    
\label{mainf}
(a) Illustration of the model. Upper: A Su-Schrieffer-Heeger chain undergoes staggered quantum jumps. Lower: The effective non-Hermitian Hamiltonian drives the short-time dynamics in the absence of the quantum-jump terms (b) The “phase diagram” of the non-Hermitian Hamiltonian \eqref{He} as $\lambda$
and $\eta$ vary. The green area represents the real spectra
gapped phase (line gapped in full spectra) and the pink area represents the real spectra
gapless phase (point gapped in full spectra), thery are seperated by the cross lines, on which
exceptional points exsits.The $\lambda=1/2$ yellow dash line corresponds
to the Hermitian case. $\lambda>1/2$($\lambda<1/2$)
corresponds to a positive (negative) imaginary part spectrum. The brown dash line represents that the model reduces to single band. Therefore the exceptional point is fake. Representative
points as denoted by different shapes on the diagram, their spectra
are shown on the right panel. In the context, we mainly focus on (III)
,(I),(IV). (c) Full spectra of distinct phases: (I),(III) and (VI) are point-gapped phases; (II) is Hermitian gapped phase; (IV) is the line gapped phase; (V) is the exceptional phase. (d) Upper: the particle number density $n_{k}(t)$ elove
with time in $k$-spce, there is left-right asymmetry. Lower: the real space particle number density $n_{x}(t)$ elove
with time under open boundary condition (OBC). It is a remanent of non-Hermitian
skin effect (NHSE), also dubbed as chiral damping effect~\cite{song_non-hermitian_2019}.  (e) Demonstration of numeric calculation of entropy. The total site number is taken as $L=200$. $S$ vs. $t$ for  (III) $\lambda=0.2,\eta=0.3$ (red); (IV) $\lambda=0.3,\eta=0.2$ (blue); (I)
$\lambda=0.4,\eta=0.5$ (green) with lattice size $L=200$.}
\end{figure*}

{\color{blue}\emph{The dissipative model.}}--
We consider a generic Hermitian dimerized chain with hopping parameters $iw$ and $iv$, where $w,v$ are real numbers, undergoing adjacent-2-site decaying quantum jumps described by $L_{n,A}=\sqrt{\ensuremath{\gamma_{A}}}(c_{n,A}-c_{n,B})$
and $L_{n,B}=\sqrt{\gamma_{B}}(c_{n,B}-c_{n+1,A})$, as shown in Fig.~\ref{mainf}(a).   For simplicity, we consider only the case with $w=(1-\eta),v=\eta$, $\gamma_{A}=(1-2\lambda)(1-\eta)$, and $\gamma_{B}=(1-2\lambda)\eta$
with $0<\lambda,\eta<1$.
The Liouvillian is thus written as 
\begin{equation}
\label{L}
\begin{array}{cl}
\mathcal{L}\rho= & \ensuremath{-\mathrm{i}[H_{\textrm{eff}},\rho]_{\textrm{nH}}+\gamma_{A}\sum_{n}(c_{n,A}-c_{n,B})\rho(c_{n,A}^{\dagger}-c_{n,B}^{\dagger})}\\
 & +\gamma_{B}\sum_{n}(c_{n+1,A}-c_{n,B})\rho(c_{n+1,A}^{\dagger}-c_{n,B}^{\dagger})
\end{array}
\end{equation},
where $\rho$ denotes the (many-body) density matrix of the system \footnote{In an open system with particle loss, the trace of time-dependent single-particle denisty
matrix is not 1 because the particle number is not conserved. Whereas, the many-body density matrix is constrained to trace 1.} and $\gamma=\gamma_{A}+\gamma_{B}$.  The non-Hermitian
commutator $-\mathrm{i}[H_{\textrm{eff}},\rho]_{\textrm{nH}}=\ensuremath{-i\left(H_{\text{eff }}\rho-\rho H_{\text{eff }}^{\dagger}\right)}$. The effective Hamiltonian is given by
\begin{equation}
\label{He}
H_{\text{eff}}=H_{\text{\text{hop}}}+H_{\text{\text{on-site}}},
\end{equation}
where
$H_{\text{\text{hop}}}=\sum_{n=1}^{N-1}(it_{2}c_{n+1,A}^{\dagger}c_{n,B}+it_{2}^{\prime}c_{n,B}^{\dagger}c_{n+1,A})+\sum_{n=1}^{N}(t_{1}c_{n,B}^{\dagger}c_{n,A}+it_{1}^{\prime}c_{n,A}^{\dagger}c_{n,B})$
is the non-reciprocal hopping part and $H_{\text{\text{on-site}}}=i\mu\sum_{n=1}^{N}(c_{n,A}^{\dagger}c_{n,A}+c_{n,B}^{\dagger}c_{n,B})$
is the on-site part with imaginary chemical potential. And the parameters
are given by $t_{1}=(1-\eta)(1-\lambda),t_{2}=\eta(1-\lambda)$; $t_{1}^{\prime}=\lambda(1-\eta),t_{2}^{\prime}=-\lambda\eta$; $\mu=(2\lambda-1)/2$. 

The effective Hamiltonian \eqref{He} gives the quasi-steady state and information of the short-time evolution of the system, while the Liouvillian \eqref{L} determines the long-time dynamics of this system. Due to its quadratic form, the Liouvillian spectrum can be obtained by diagonalizing $\mathcal{L}$ with the so-called third quantization \cite{Prosen_2008} as $\mathcal{L}=\sum_{\underline{\nu}} \lambda_{\underline{\nu}} \bar{\eta}_{\underline{\nu}}^{\dagger} \eta_{\underline{\nu}}$, where $\bar{\eta}_{\underline{\nu}}^{\dagger}, \eta_{\underline{\nu}}$ are fermionic operators and $\lambda_{\underline{\nu}}$ denotes the eigenvalues of excitations above the Fock vacuum of $\eta_{\underline{\nu}}$. The Liouvillian superoperator acts on the (many-body) density matrix, therefore its full spectrum is a many-body version of eigenspectrum generated from single-particle eigenvalues $\lambda_{\underline{\nu}}$. 

For a non-interacting system, all the information has been encoded in the two-point correlation functions, which are defined as,
\begin{equation}
\mathcal{C}_{ij}(t)=\text{Tr}\left[c_{i}^{\dagger}c_{j}\rho(t)\right]
\end{equation}
where $i$, $j$ are index of lattice sites.
The temporal evolution of the correlation function is determined by $\dot{\mathcal{C}}_{ij}(t)=\textrm{Tr}\left[c_{i}^{\dagger}c_{j}\dot{\rho}(t)\right]$. Combined with the Lindblad master equation \eqref{M} \footnote{The steady-state is the Hilbert vacuum with $\mathcal{C}(\infty)$=0},
the time-dependent correlation function in matrix form is given by \cite{supp}
\begin{equation}
\label{Ct}
\mathcal{C}(t)=e^{i\mathcal{D}t}\mathcal{C}(0)e^{-i\mathcal{D}t}
\end{equation}
where $\mathcal{D}=H_{0}^{T}+i\sum_{i}L_{l}^{\dagger}L_{l}/2$ is called the damping operator. The eigenvalues $\lambda_{\mathcal{D}}$ of $\mathcal{D}$ are related to the Liouvillian
spectrum $\lambda_{\underline{v}}$ by $\lambda_{\underline{v}}=-i\lambda_{\mathcal{D}}$ under
purely loss or gain\cite{supp}.

The Hamiltonian is easy to diagonalize in the $k$-space. We can get the energy spectrum of this model in periodic
boundary condition (PBC), $\epsilon_{k}=\pm\left[-(t_{2}\prime e^{ik}+t_{1})(t_{2}e^{-ik}+t_{1}^{\prime})\right]^{1/2}+i\mu$
as shown in Fig.~\ref{mainf}(c). The point-gap and line-gap topology is discussed in literature \cite{lee_anomalous_2016,kawabata_classification_2019,leykam_edge_2017,yao_edge_2018,ding_emergence_2016,bergholtz_exceptional_2021,helbig_generalized_2020,luo_higher-order_2019,yang_jones_2020,borgnia_non-hermitian_2020,xue_non-hermitian_2020,song_non-hermitian_2019-1,song_p_2020,zhou_periodic_2019,liu_second-order_2019,kawabata_topological_2019,liang_topological_2013,sato_time-reversal_2012}. The famous non-Hermitian skin effect (NHSE) \cite{yao_edge_2018,song_non-hermitian_2019,okuma_topological_2020} occurs in OBC instead. However, in PBC, the NHSE corresponds to the left-right asymmetry in momentum-space, which will be discussed later.
The system has been classified as line gapped phases and point gapped phases\footnote{On the diagonal crossline ($\lambda(1-\lambda)=\eta(1-\eta)$), the energy spectrum cross the
exceptional point (EP) in moment space, which have exotic states coalesce
and defective behaviors. More details on exceptional points can be found in literature\cite{berry_physics_2004,heiss_physics_2012,bergholtz_exceptional_2021,yi2023exceptional}}, as shown in Fig.~\ref{mainf}(b), according to the gap structure in the real part of the spectrum of the effective Hamiltonian.
The ground state of the system is constructed by filling all the single particle states with negative real energy. Instead, 
the gap in Liouvillian spectrum is related to the spectrum of the Hamiltonian through $\Lambda=-\min\{\textrm{Im[\ensuremath{\epsilon_{n}}]}\}$. And 
whether the Liouvillian spectrum is gapped or not determines mainly the dyamical behaviors\cite{cai_algebraic_2013}.
 The gap of the Liouvillian spectrum closes only when $\eta=1/2$.



{\color{blue}\emph{Entropy evolution in real space.}}--
Now we consider both the short-time and long-time evolution of the system. Through Eq.~\eqref{Ct}, one can easily calculate the time-dependence of a single-particle operator, for example, the particle number density, of which the result is shown in Fig.~\ref{mainf}(d). For a better understanding of the result, we consider
a heuristic example of a two-site system with distinct initial states. 

\begin{figure}
\includegraphics[scale=0.33]{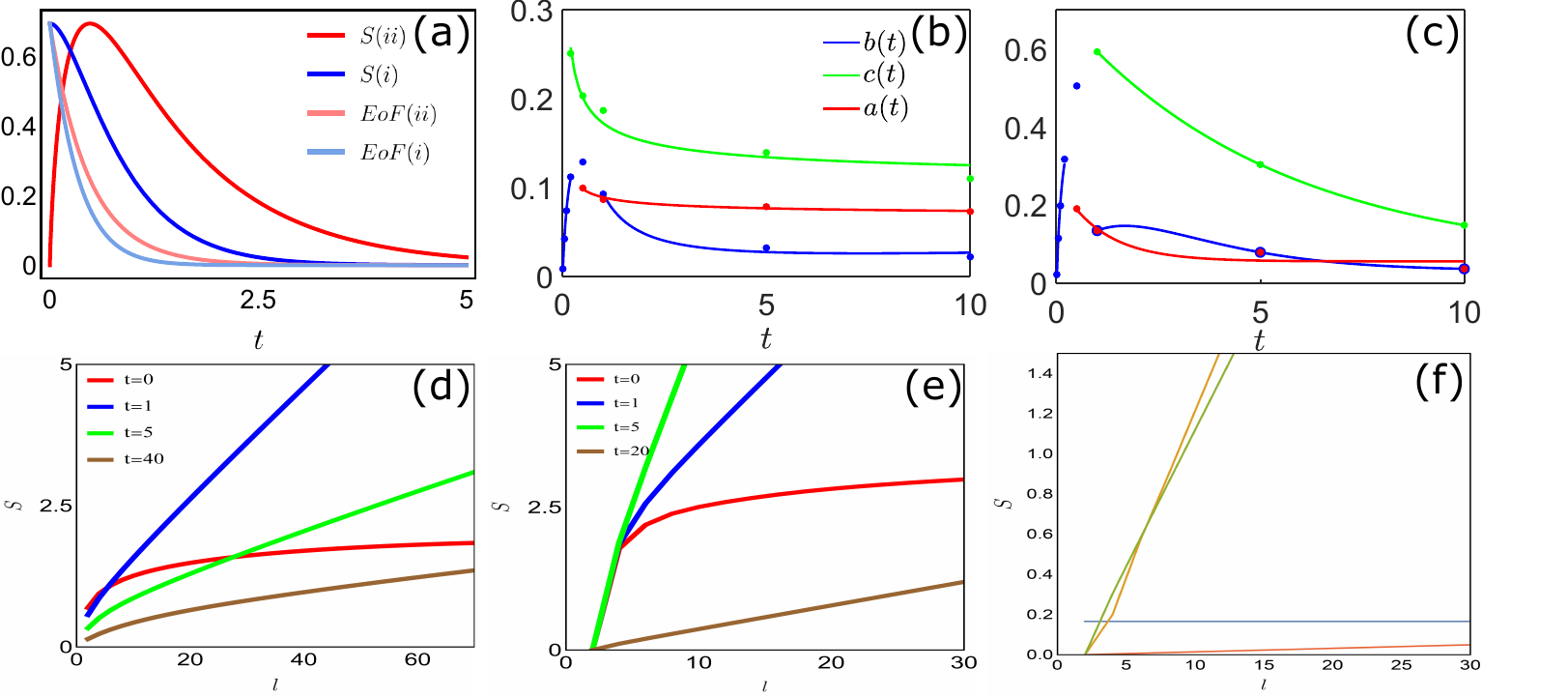}

\caption{\label{En}(a) Evolution of entropies and entanglement of formation (EoF) with time for two different initial states $(i)$ and $(ii)$. The short-time increase of $S(ii)$ is $\sim-\gamma t\ln(\gamma t)$ and the long-time decay is $\gamma t\exp(-2\gamma t)$ (For $S(i)$, it is $\gamma t\exp(-\gamma t)$). EoF$(i)$ is $\sim \gamma t\exp(-2\gamma t)$ and for $S(ii)$, it is $\sim \gamma t\exp(-4\gamma t)$. (b) Fitting of entropy using $S(t)\sim a(t)\ln(l)+b(t)l+c(t)$ for gapped Liouvillian and gapless Hamiltonian. For short time, $b(t)\sim -1.12 t \ln(t) -0.03$, and for long time, $b(t)\sim (0.06 t+0.11) \exp(-0.39 t)+0.002$, $a(t)\sim(0.01 t+0.16) \exp(-0.41 t)+0.004$, and $c(t)\sim(0.07 t+0.61) \exp(-0.38 t)+0.008$. (c) For gapless Liouvillian and gapless Hamiltonian, $b(t)\sim -0.38 t \ln(t) -0.01$ for short time, and for long time, $b(t)\sim(0.10 +0.02 \ln(t)) t^{-0.51}+0.001$, $a(t)\sim(0.10 +0.006 \ln(t)) t^{-0.45}+0.007$, and $c(t)\sim(0.12 +0.02 \ln(t)) t^{-0.49}+0.006$. (d)-(f) Demonstration of numeric calculation of entropy. The total site number is taken as $L=200$. (d) $S$ vs. $l$ of the $\lambda=0.4,\eta=0.5$ case in various time scales. The top right corner shows the $S=\alpha\ln\left[\sin\left(\pi l/L\right)\right]$ fitting at $t=0$, with $\alpha=0.333$. (e) $S$ vs. $l$ of the $\lambda=0.2,\eta=0.3$ case in various time scales. The top right corner shows the $S=\alpha\ln\left[\sin\left(\pi l/L\right)\right]$ fitting at $t=0$ with $\alpha=0.333$. (f) $S$ vs. $l$ of the $\lambda=0.2,\eta=0.3$ case in various time scales. At $t=0$, the entropy is constant.}

\end{figure}

We examine the minimal system with 3 distinct initial states as a heuristic example to understand the time evolution:
$(i)$ maximally entangled state $|\psi(0)\rangle_{AB}=\sqrt{\frac{1}{2}}|10\rangle_{AB}-\sqrt{\frac{1}{2}}|01\rangle_{AB}$, which is a ground state of the 2-site model in Eq.~\eqref{He}; $(ii)$ unentangled excited state $|11\rangle$; $(iii)$ the dark state $|00\rangle$, i.e., without time-evolution, also the final state. The time-dependent density matrix, as well as the reduced density matrix and entanglement entropy $S_{A(B)}=-\text{Tr}\rho_{A(B)}\text{ln\ensuremath{\rho_{A(B)}}}$, can be analytically obtained from the 2-site Lindblad master equation. The results are plotted in Fig.~\ref{En}(a).

For a pure state, the quantum entanglement between bipartite parts A and B can be measured by the entanglement entropy .
For a thermalized mixed state, classical information is included in entropy. Thus, one can adopt the entanglement of formation (EoF) \cite{hill_entanglement_1997,wootters_entanglement_1998},
$\mathcal{E}_{F}\left(\rho_{\mathrm{mixed}}\right)\eqqcolon\min\sum_{i}p_{i}S\left(\left|\psi_{\mathrm{pure}}^{i}\right\rangle \right)$ to measure the 
pure quantum entanglement, which counts the minimum number of singlets in demand to create an ensemble of pure states since singlets are basic unit of entanglement. The minimum is over all possible pure-state decompositions of
$\rho_{\mathrm{mixed}}$ with probabilities $p_{i}$. Through
two-qubit concurrence $\mathfrak{C}(\rho)=\max\left\{ 0,\lambda_{1}-\lambda_{2}-\lambda_{3}-\lambda_{4}\right\} $,
where $\lambda_{i}$ are square roots of the eigenvalues of the 
spin-flipped state matrix in decreasing order, The EoF is given
by\cite{hill_entanglement_1997,wootters_entanglement_1998} 

\begin{align*}
      \mathcal{E}_{F}(\rho)=h\left(\frac{1+\sqrt{1-\mathfrak{C}^{2}}}{2}\right),\\
     h(x)=-x\log_{2}x-(1-x)\log_{2}(1-x).
\end{align*}

 As shown in Fig.~\ref{En}(a), for short-time dynamics,
the classical information proliferates as a manifestation of rapid
thermalization, in which the entropy can capture the physics while EoF ignores it. While the entropy perfectly describe the decay of entanglement in long time regime.\footnote{We obtain $\mathfrak{C}(t)=e^{-4\gamma t}$, which coincides with the
non-trivial eigenvalues of $\rho_{t}$. The time-dependent EoF decreases monotonically as $\mathcal{E}_{F}(t)\propto\gamma t\text{exp}(-8\gamma t)$. }
 This suggests that entropy can be regarded as a reliable quantity for describing the dynamical behavior of the system throughout their entire temporal evolution.

Before encountering a true many-body problem, we can gain insight from the results obtained for two-site systems at the microscopic level. The ground state is a linear combination of three distinct types of initial states labeled $(i)$, $(ii)$, and $(iii)$, as mentioned above. When the ground state is gapped, the area law dictates that entanglement is only present in the vicinity of the cut, contributed by the $O(l^0)$ $(i)$-states. However, thermalization excites a number of $O(l)$ $(ii)$-states, resulting in a total entropy evolution behavior that is a combination of $(i)$ and $(ii)$ . In the case of gapless systems, there are $O(\text{ln}l)$ $(i)$-states that contribute to the initial entropy instead of $O(l^0)$, while the thermalization process remains almost the same.
In the presence of a Liouvillian gap, the particle density decay rate is dominated by the exponential factor $\exp-\Lambda t$. For gapless systems, however, the collective decay rate of the density contributes a factor of $\int d(\delta k) \exp \left[-(\delta k)^2 t\right] \sim t^{-1 / 2}$ in our study case. As a result, the entropy evolution during the evaporization process is distinguished by whether the Liouvillian gap closes or not.

To calculate the entanglement entropy of a many-body system with spatial bipartition, we employ the correlation matrix method developed in Refs. \cite{peschel_calculation_2003,peschel_reduced_2009,chung_density-matrix_2001}. Despite the complexity of the problem, the Wick theorem still holds \cite{herviou_entanglement_2019,chen_entanglement_2021,maity_growth_2020,sharma_landauer_2015}. Owing to the quadratic form of the Liouvillian, we can describe the problem using a single-particle picture, noting that the single-particle density matrix $\tilde{\rho}_{t}$ is not trace-normalized to 1. The entanglement Hamiltonian can still be written as a quadratic form and can therefore be simultaneously diagonalized. Thus, the reduced (biorthogonal \cite{brody_biorthogonal_2013}) density matrix can be expressed as $\tilde{\rho}_{t}=\text{exp}\left(\sum_{\alpha}\varepsilon_{\alpha}(t)a_{\alpha}^{\dagger}a_{\alpha}\right)/\mathcal{Z}_{0}$, where $\mathcal{Z}_{0}$ is the partition function and the normalization factor, and it satisfies $\text{Tr}\tilde{\rho}_{0}=1$. The particle loss in real space is reflected by the vanishing non-trivial mid-gap entanglement energies $\varepsilon_{\alpha}$. We initialize the ground state of the system at $t=0$ with half-filling of the real energy band. The entropy
is expressed by
\begin{equation}
\begin{split}S(t)= & -\sum_{\sigma}[C_{\sigma}(t)\ln C_{\sigma}(t)-C_{\sigma}(t)\ln\left(\mathcal{Z}_{t}/\mathcal{Z}_{0}-C_{\sigma}(t)\right)\\
 & +\mathcal{Z}_{t}/\mathcal{Z}_{0}\ln\left(1-C_{\sigma}(0)\right)]
\end{split}
\end{equation}
where $C_{\sigma}(t)$ are the eigenvalues of the reduced correlation
matrix $\tilde{\mathcal{C}}(t)$, and $\mathcal{Z}_{t}=\mathcal{Z}_{0}\text{Tr}\tilde{\rho}_{t}$.

We calculate numerically the entropy with a spatial partition
of the observed system with total size $L$ to part A with length $l$ and part B with length $L-l$
. At $t=0$, the entropy $S$ in the gapped phases is equal to a size-independent constant $a_0$, as shown in Fig.~\ref{En}(d), which is in accordance with the area law. For the gapless phase, the entropy shown in Fig.~\ref{En}(b)(c) fits as $S=0.333 \ln\left[\sin\left(\pi l/L\right)\right]+c_{0}$, which corresponds to a conformal field theory (CFT) with central charge $c=1$\cite{holzhey_geometric_1994,calabrese_entanglement_2009}. 
The results above can be analytically obtain via Toeplitz determinants\cite{jin_quantum_2004}. The absence of any $l$-linear dependence of the entanglement entropy is a consequence of two point-like Fermi surfaces in 1d\cite{PhysRevB.105.205403}. The logarithmic scaling observed in this case is ascribed to the discontinuities of occupied states on these Fermi points.
However, when the quantum jumps are turned on, the system undergoes thermalization, and the Fermi surface is no longer well-defined as time evolves. This results in the appearance of a $l$-linear-dependent entropy as the system evolves.

We then compute the temporal evolution of the entropy $S$ and the results are presented in Tab.~\ref{tab:cases} and Fig.~\ref{En}, which are consistent with our expectations and numeric fitting ~\footnote{Analytical calculation of asymptotic behavior of the determinants of Toeplitz matrices implies $a(t)\sim\text{Li}_{2}\rho_{t}\sim\rho_{t}\sim n(t)$, when t is large and $\rho_t\ll1$.}\textsuperscript{,}~\footnote{Based on the intuition gained from the results of two qubits and numerical fitting, we have observed that the long-term behavior of the system exhibits a higher-order leading term in time, as compared to the analytical expectation. However, it is worth noting that the coefficient associated with the $t$-leading terms is significantly smaller than that of the sub-leading terms. Consequently, the reduction in entropy resulting from the particle loss (primarily classical, but dominant during the intermediate time period) cannot be attributed solely to this process, but also arises from decoherence, as evidenced by the results observed in the two-qubit case.}.  Notably, the time-dependent entropy in Fig.~\ref{mainf}(e) exhibits the same growth behavior in the short time limit, which can be described by the function $S(t)\sim-t\text{ln}t$. This formula also coincides with the expression for the short-time entropy increase in one-dimensional dissipative interacting fermions reported in Ref.\cite{bacsi_vaporization_2020}. We emphasize that this behavior of entropy growth in the short time limit is ubiquitous in one-dimensional dissipative fermion liquids.

\begin{table*}
\begin{centering}
\begin{tabular}{c|cccc}
\hline 
\multicolumn{1}{c|}{Liouvillian} & \multicolumn{2}{c}{gapped} & \multicolumn{2}{c}{gapless}\tabularnewline
\hline 
\hline 
Hamiltonian & \multirow{2}{*}{thermalization} & \multirow{2}{*}{evaporization} & \multirow{2}{*}{thermalization} & \multirow{2}{*}{evaporization}\tabularnewline
(real part) &  &  &  & \tabularnewline
\hline 
\multirow{2}{*}{gapped} & $[O(l)-O(l^{0})]$$\times$ & $[O(l)+O(l^{0})]$$\times$ & $[O(l)-O(l^{0})]$$\times$ & $[O(l)+O(l^{0})]$$\times$\tabularnewline
 & $-t\text{ln}t+O(t)$ & $(t+O(t^{0}))\exp(-\Lambda t)$ & $-t\text{ln}t+O(t)$ & $t^{-1/2}(\text{ln}t+O(t^{0}))$\tabularnewline
\hline 
\multirow{2}{*}{gapless} & $[O(l)-O(\text{ln}l)]$$\times$ & $[O(l)+O(\text{ln}l)]$$\times$ & $[O(l)-O(\text{ln}l)]$$\times$ & $[O(l)+O(\text{ln}l)]$$\times$\tabularnewline
 & $-t\text{ln}t+O(t)$ & $(t+O(t^{0}))\exp(-\Lambda t)$ & $-t\text{ln}t+O(t)$ & $t^{-1/2}(\text{ln}t+O(t^{0}))$\tabularnewline
\hline 
\end{tabular}
\end{centering}
\caption{\label{tab:cases}Evolution of entropy with time (showing leading and sub-leading terms) in different cases of Hamiltonian and Liouvillian spectra. }

\end{table*}


{\color{blue}\emph{Correlation in Momentum Space.}}--
Upon Fourier transforming the correlator given by Eq.\eqref{Ct} into momentum space, the dynamics of quasi-particles in k-space can be readily obtained, as illustrated in Fig.\ref{mainf}(d). This plot exhibits a left-right asymmetry, wherein the sign of the group velocity $v=dE/dk$ identifies the left-moving (right-moving) particle, such that $\text{sign}[v]=-1(+1)$ corresponds to left-movers (right-movers). This asymmetry is a consequence of the non-reciprocal hoppings and is indicative of NHSE in OBC since particles tend to accumulate at one side if there are endpoints.

We now turn our attention to the time evolution of a system subject to both density-density interaction and particle loss. Let us assume that the particle loss is activated 
 at $t=0$. At short times, the Fermi surface remains well-defined, and consequently, the short-time dynamics can be examined using the Luttinger liquid Hamiltonian via bosonization:
\begin{equation}
H=\sum_{q\neq0}v|q|b_{q}^{\dagger}b_{q}+\frac{g_{2}|q|}{2}\left[b_{q}b_{-q}+b_{q}^{+}b_{-q}^{+}\right],
\end{equation}
where $v|q|$ represents the spectrum with velocity v, $b_{q}^{\dagger}=i\sqrt{1/|q|}\sum:c_{k+q}^{\dagger}c_{k}:$
create electron-hole pairs around the Fermi surface, and the parameter $g_{2}$ denotes the strength of the interaction. After bosonization, the quantum jump operator becomes an exponential of a linear combination of boson operators. As the Hamiltonian is quadratic, the non-unitary dynamics of $\rho$ is encoded in the momentum-space correlation function $\mathcal{C}_{k+q,k}(t)=\text{Tr}\left[c_{k+q}^{\dagger}c_{k}\rho(t)\right]$. With Eq.~\eqref{L}, the equation
of motion of $b_{q}^{\dagger}(t)$ is given by 

    \begin{align}
   \partial_{t}b_{q}^{\dagger}=-4\gamma b_{q}^{\dagger}+iv|q|b_{q}^{\dagger}-ig_{2}|q|b_{-q},\\
\partial_{t}b_{-q}=-iv|q|b_{-q}-ig_{2}|q|b_{q}^{\dagger}.   
    \end{align}

  Despite the complexity of obtaining a full solution, we can focus on short-time dynamics to express $b_{q}^{\dagger}(t)$ and $b_{-q}(t)$ as $b_{q}^{+}(t)=\exp(iv|q|t-4\gamma t)b_{q}^{+}-ig_{2}|q|tb_{-q}$ and $b_{-q}(t)=\exp(-iv|q|t)(1-ig_{2}|q|t)b_{-q}-\dfrac{1}{2}g_{2}^{2}|q|^{2}t^{2}b_{q}^{+}$ by ignoring high-order terms. However, it is not generally possible to equally bipartite the system into two parts ($q>0$ and $q<0$) in momentum space. This is because the $q>0$ part decays exponentially, while the particle numbers at $q<0$ remain nearly unchanged.

Moreover, the momentum-space entanglement entropy (MSEE) should exhibit an overall decay factor as $\exp(-8\gamma t)$ for the summation of $q$. Nonetheless, if the time $t$ is short enough such that the loss event does not happen, the density matrix is simply renormalized, and the canonical commutation relation of $\left[b_{q}(t),b_{q}^{+}(t)\right]=1$ can be satisfied by setting $b_{q}^{+}(t)=u_{q}(t)b_{q}^{+}+v_{q}(t)b_{-q}$ \footnote{Ignoring the nonlinear $b_{q}^{+},b_{q}$ terms and $b_{-q}(t)=u_{q}^{*}(t)b_{-q}-v_{q}(t)b_{q}^{+}$. It is reasonable to assume $\left|u_{q}(t)\right|^{2}+\left|v_{q}(t)\right|^{2}=1$ as in the $\mathcal{PT}$-symmetric non-Hermitian quench in Ref.\cite{dora_quantum_2020} }. 
We can calculate the short-time pseudo-Bogoliubov coefficients as $u_{q}(t)\sim1+iv|q|t-4\gamma t$ and $v_{q}(t)\sim-2\sqrt{2}i\gamma^{1/2}t^{1/2}$, with the $g_{2}$ dependence appearing only in higher-order terms. As a result, the MSEE increases with time as $S\sim-t\ln t$ ($t\ll1$), suggesting an interaction-independent $q$,$-q$ entanglement that originates from the coherent nature of $L_{q}$. This result reflects the correlation effect of the asymmetric hopping, which has the same origin as NHSE.

{\color{blue}\emph{Discussion.}}--
We investigated the entropy of a one-dimensional dynamical quantum liquid in various scenarios. Although the entropy provides information beyond quantum entanglement, it offers a more comprehensive representation of the universal behavior of the system's temporal evolution. While the logarithmic entanglement negativity \cite{PhysRevLett.77.1413,PhysRevLett.95.090503} is a viable method for analyzing the pure quantum entanglement spectrum, its computational complexity is significant. The manipulation of cold atomic optical lattice systems allows for the experimental engineering of many-body open quantum systems, where atomic couplings and parallel inciding lasers can be fine-tuned \cite{zhou_rapid_2021}. Alternatively, the many-body Liouvillian dynamics can be simulated on large-scale dissipative three-level superconducting qubit circuits by controlling the driven frequencies and couplings between qubits. The non-Hermitian ground state is prepared by quantum tomography \cite{naghiloo_quantum_2019}, and entanglement can be measured using several protocols \cite{abanin_measuring_2012,daley_measuring_2012,choo_measurement_2018,dalmonte_quantum_2018,tran2023measuring}.Additionally, other numerical or qubit-based quantum simulation techniques, e.g. in \cite{varma_simulating_2021,weimer_simulation_2021}, are also feasible for open systems.

{\color{blue}\emph{Acknowledgement.}}--
We acknowledge Yong-Shi Wu for inspiring comments at the early stage of this work. We also acknowledge Gregory A. Fiete for fruitful discussion. This work was supported by NSFC (Grants No. 11861161001), the Science, Technology and Innovation Commission of Shenzhen Municipality (No. ZDSYS20190902092905285), Guangdong Basic and Applied Basic Research Foundation under Grant No. 2020B1515120100, Shenzhen-Hong Kong Cooperation Zone for Technology and Innovation (Grant No.HZQB-KCZYB-2020050), and Center for Computational Science and Engineering at Southern University of Science and Technology.

\bibliography{main}
\end{document}